\documentclass[11pt]{article}

\usepackage{amsmath, amssymb}
\usepackage{enumerate}
\numberwithin{equation}{section}

\allowdisplaybreaks

\usepackage{graphicx}
\newcommand{\p}{\partial}
\newcommand{\e}{\epsilon}

\newcommand{\CC}{\mathbb{C}}

\newcommand{\I}{\mathcal{I}}
\newcommand{\J}{\mathcal{J}}
\newcommand{\F}{\mathcal{F}}
\newcommand{\D}{\mathcal{D}}

\renewcommand{\L}{\mathcal{L}}

\newcommand{\beq}{\begin{equation}}
\newcommand{\eeq}{\end{equation}}

\newcommand{\nn}{\nonumber}

\newtheorem{dfn}{Definition}[section]
\newtheorem{lem}[dfn]{Lemma}

\newtheorem{thm}[dfn]{Theorem}

\newtheorem{cor}[dfn]{Corollary}
\newtheorem{emp}[dfn]{Example}

\newenvironment{prf}{\noindent {\it Proof} \ }{\hfill $\Box$}
\newenvironment{prfn}[1]{\noindent {\it Proof of #1} \ }{\hfill $\Box$}

\begin{document}

\title{The loop equation for the Burgers--KdV hierarchy}
\author{Di Yang,\quad Chunhui Zhou}
\date{}
\maketitle

\begin{abstract}
The Burgers--KdV hierarchy was introduced towards understanding  
intersection numbers on the moduli space of Riemann surfaces with boundaries \cite{Bu2, PST}. 
The goal of this paper is to derive the Dubrovin--Zhang type loop equation 
for the topological tau-function 
of the Burgers--KdV hierarchy. 
As as application, we provide some relations for open intersection numbers.
\end{abstract}

%\tableofcontents

\section{Introduction}

Let $w,\rho$ be indeterminates, and let $(\mathcal A=\mathbb C[w,\rho,w_x,\rho_x,\dots], \p=\p_x)$ be the differential algebra. 
Recall that the Burgers--KdV hierarchy is the following integrable hierarchy introduced by Buryak~\cite{Bu2} (cf.~\cite{PST}):
\begin{align}
&\frac{\p w}{\p t_n}=\p_x K_n, \quad \frac{\p \rho}{\p t_n}=\p_x R_n, \quad n\geq0,\label{openkdv1}\\ 
&\frac{\p w}{\p s_n}=0,  \quad\quad\,\,\,\,\,  \frac{\p \rho}{\p s_n}=\p_x Q_n,\quad n\geq0,\label{openkdv2}
\end{align}
where 
$K_n, R_n, Q_n\in \mathcal{A}[\e]$
are given by
\begin{align}
&K_0=w,\quad R_0=\rho,\quad Q_0=w+\frac{\rho^2}{2}+\e \frac{\rho_x}{2}, \label{K0Q0R0}\\
&\p_x K_n=\frac{2}{2n+1}\Big(w \p_x+\frac12 w_x+\frac{\e^2}{8}\p_x^3\Big)K_{n-1}, 
\label{defK}\\
&R_n=\frac{2}{2n+1}\bigg(\Big(w+\frac{\rho^2}{2}+\e\big(\frac{\rho_x}{2}+\rho\p_x\big)+\frac{\e^2}2\p_x^2\Big) R_{n-1}+\Big(\frac\rho2+\frac{3}4 \e\p_x\Big)K_{n-1}\bigg),
\label{defR}\\
&Q_n=\frac1{n+1}\Big(w+\frac{\rho^2}{2}+\e\big(\frac{\rho_x}{2}+\rho\p_x\big)+\frac{\e^2}2\p_x^2\Big) Q_{n-1},\quad n\geq1. \label{defQ}
\end{align}
Here the integration constants in \eqref{defK} are chosen as zero.
Equations~\eqref{openkdv1} with $n=0$ read
\beq
\frac{\p w}{\p t_0}=\frac{\p w}{\p x},\quad \frac{\p \rho}{\p t_0}=\frac{\p \rho}{\p x},
\eeq
so we identify $t_0$ with $x$.

For an arbitrary solution 
$(w,\rho)$ in $\CC[[{\bf t};\e]]\times\CC[[{\bf t},{\bf s};\e]]$ 
to the Burgers--KdV hierarchy, 
there exists a tuple 
$(\tau_1,\tau_2)\in \CC((\e))[[{\bf t}]]\times\CC((\e))[[{\bf t},{\bf s}]]$
satisfying 
\beq\label{wrhotau}
w=\e^2\p_x^2\log\tau_1,\quad \rho=\e\p_x\log\tau_2
\eeq
and a sequence of similar conditions (see Section~\ref{sec2} for the details), 
called~\cite{YZ} the tau-tuple of the solution $(w,\rho)$. 
We call the product $\tau=\tau_1\tau_2$ the tau-function of the solution.

We are particularly interested in the solution $(w_{\rm top},\rho_{\rm top})\in \CC[[{\bf t};\e]]\times \CC[[{\bf t},{\bf s};\e]]$ specified by
\beq
(w_{\rm top},\rho_{\rm top})|_{t_n=x\delta_{n,0},s_n=0,n\geq0}=(x,0).
\eeq
We call $(w_{\rm top},\rho_{\rm top})$ the {\it topological solution} to the Burgers--KdV hierarchy.
The tau-function $\tau_{\rm top}=\tau_{1,\rm top}\tau_{2,\rm top}$ of this solution 
can~\cite{Bu2} be further chosen to satisfy the equation
\beq\label{string}
\sum_{n\geq1}\Big(\tilde t_n \frac{\p \tau_{\rm top}}{\p t_{n-1}}+s_n\frac{\p\tau_{\rm top}}{\p s_{n-1}}\Big)
+\frac{t_0^2}{2\e^2}+\frac{s_0}{\e}=0,
\eeq
where $\tilde t_n=t_n-\delta_{n,1}$.
In this way, $\tau_{\rm top}$ is unique up to a constant factor.
We call $\tau_{\rm top}$ the {\it topological tau-function} of the Burgers--KdV hierarchy.

According to the Witten--Kontsevich theorem \cite{K, W}, the first component $\tau_{1,\rm top}$ has the form
\beq
\tau_{1,\rm top}({\bf t};\e)=e^{\sum_{g\geq0}\e^{2g-2}\F^c_g({\bf t})},
\eeq
where $\mathcal F_g^c({\bf t})$ has the expression
\beq
\mathcal F_g^c({\bf t})=\sum_{n\geq0}\sum_{k_1,\dots,k_n\geq0}\frac{\langle\tau_{k_1}\cdots\tau_{k_n}\rangle_g^c}{n!}t_{k_1}\cdots t_{k_n}, \quad 
\langle\tau_{k_1}\cdots\tau_{k_n}\rangle_g^c=\int_{\overline{\mathcal M}_{g,n}}\psi_1^{k_1}\cdots \psi_n^{k_n},
\eeq
where $\overline{\mathcal M}_{g,n}$ denotes the moduli space of smooth complex algebraic curves of genus $g$ with $n$ distinct marked points,
and $\psi_i$ denotes the first Chern class of the tautological line bundle on~$\overline{\mathcal M}_{g,n}$.
We call $\mathcal{F}_g^c({\bf t})$ the genus $g$ topological free energy for the KdV hierarchy.
It is well-known that $\tau_{1,\rm top}$ satisfies the following Virasoro constraints:
\beq\label{VirasoroKdV}
\mathcal L_m \tau_{1,\rm top}=0,\quad m\geq-1.
\eeq
where $\mathcal L_m$ are operators as follows:
\begin{align}
\mathcal L_{-1}
=&\sum_{n\geq1}\tilde t_n \frac{\p }{\p t_{n-1}}+\frac{t_0^2}{2\e^2}, \\
\mathcal L_m
=&\sum_{n\geq0}\frac{(2n+2m+1)!!}{2^{m+1}(2n-1)!!}\tilde t_n \frac{\p}{\p t_{n+m}} 
+\frac1{16}\delta_{m,0} \nn \\
&+\frac{\e^2}{2}\sum_{n=0}^{m-1}\frac{(2n+1)!!(2m-2n-1)!!}{2^{m+1}}\frac{\p^2}{\p t_n\p t_{m-n-1}},
\quad m\geq0.
\end{align}

The second component has the form
\beq
\tau_{2,{\rm top}}({\bf t},{\bf s};\e)=e^{\sum_{p\geq0}\e^{p-1}\F^o_p({\bf t},{\bf s})},
\eeq
where $\mathcal F_p^o({\bf t},{\bf s})$ has the expression 
\beq\label{Fogts}
\F^o_p({\bf t},{\bf s})=\sum_{m,n\geq0}\sum_{\substack{k_1,\dots,k_m\geq0,\\\ell_1,\dots,\ell_n \geq0}}
\frac{\langle\tau_{k_1}\cdots\tau_{k_m}\sigma_{\ell_1}\cdots\sigma_{\ell_n}\rangle_p^o}{m!n!}t_{k_1}\dots t_{k_m} s_{\ell_1}\dots s_{\ell_n},
\eeq
where the correlators
$\langle\tau_{k_1}\cdots\tau_{k_m}\sigma_{\ell_1}\cdots\sigma_{\ell_n}\rangle_p^o$
are (cf.~\cite{Bu1, Bu2,PST}) related with intersection numbers on moduli space of Riemann surfaces with boundaries. 
We call $\mathcal F_p^o({\bf t},{\bf s})$ the genus $p$ topological open free energy for the Burgers--KdV hierarchy.

Recall from e.g. \cite{DY, DZ-norm, EYY} that for every $g\ge1$
there exists a function of $(3g-2)$ variables
\[
F^c_g=F^c_g(v_1,v_2,\dots,v_{3g-2})
\]
such that 
\beq\label{closedjet}
\F_g^c({\bf t})=F^c_g\bigg(\frac{\p v_{\rm top}({\bf t})}{\p t_0},\dots,\frac{\p^{3g-2}v_{\rm top}({\bf t})}{\p t_0^{3g-2}}\bigg),
\eeq
where $v_{\rm top}({\bf t}):=\p_{t_0}^2\F^c_0({\bf t})$.
For example,
\beq\label{Fc12}
F^c_1=\frac1{24}\log v_1,\quad F^c_2=\frac{v_4}{1152v_1^2}-\frac{7v_2v_3}{1920v_1^3}+\frac{v_2^3}{360v_1^4}.
\eeq

By using a method from~\cite{DZ-norm} (see also \cite{LYZZ}), we will show in Section~\ref{sec2} the following theorem,
which gives the jet-variable representations for the topological open free energies $\F^o_p$, $p\ge1$.
\begin{thm}\label{main0}
There exists a sequence of functions 
\begin{align}
&F_1^o=\frac1{2}\log(v_1+rr_1), \label{Fo1} \\
&F^o_p=F^o_p(r,v_1,r_1,\dots,v_{2p-1},r_{2p-1}) \nn  \\
&\qquad \in 
v_1^{-(5p-2)}(v_1+rr_1)^{-(5p-2)}\CC[r, v_1,r_1,v_2,r_2,\dots,v_{2p-1},r_{2p-1}],
\quad p\geq2 \label{Fog}
\end{align}
satisfying the homogeneous conditions
\begin{align}
&\sum_{k\geq 1}k v_k \frac{\p F_p^o}{\p v_k}+\sum_{k\geq0}k r_k \frac{\p F_p^o}{\p r_k}=(p-1)F_p^o, \quad p\geq2
\label{homogeneous1}
\\
&\sum_{k\geq 1}(k-1) v_k \frac{\p F_p^o}{\p v_k}+\sum_{k\geq0}\left(k-\frac12\right) r_k \frac{\p F_p^o}{\p r_k}=\frac32(p-1)F_p^o,\quad p\geq2
\label{homogeneous2}
\end{align}
such that for $p\geq1$,
\beq
\F^o_p({\bf t},{\bf s})=F^o_p\bigg(r_{\rm top}({\bf t},{\bf s}), \frac{\p v_{\rm top}({\bf t})}{\p t_0},\frac{\p r_{\rm top}({\bf t},{\bf s})}{\p t_0},\dots, \frac{\p^{2p-1} v_{\rm top}({\bf t})}{\p t_0^{2p-1}},\frac{\p^{2p-1}r_{\rm top}({\bf t},{\bf s})}{\p t_0^{2p-1}}\bigg),
\eeq
where $r_{\rm top}({\bf t},{\bf s})=\p_{t_0}\F^o_0({\bf t},{\bf s})$.
\end{thm}

It was shown in~\cite{Bu2} that the topological tau-function $\tau_{\rm top}$ 
satisfies the following Virasoro constraints:
\beq\label{Virasoroconstraints}
\mathcal L_m^{\rm ext}\tau_{\rm top}=0,\quad m\geq-1,
\eeq
where $\mathcal L^{\rm ext}_m$ are operators given by
\begin{align}
\mathcal L_{-1}^{\rm ext}=&\mathcal L_{-1}+\sum_{n\geq1}s_n\frac{\p}{\p s_{n-1}}+\frac{s_0}{\e}, \label{Lminus1} \\
\mathcal L_m^{\rm ext}
=&\mathcal L_m+
%\sum_{n\geq0}\frac{(2n+2m+1)!!}{2^{m+1}(2n-1)!!}\tilde t_n \frac{\p}{\p t_{n+m}} 
\sum_{n\geq0}\frac{(n+m+1)!}{n!}s_n\frac{\p}{\p s_{n+m}}
+\frac{3(m+1)!}{4}\e\frac{\p}{\p s_{m-1}}+\frac34\delta_{m,0}
\quad m\geq0. \label{openVirasoro}
\end{align}

Introduce the following notations:
\beq\label{defab}
A=\frac{1}{\lambda-v},\quad B=\frac{1}{\lambda-v-\frac{r^2}2}, \quad
\p=\sum_{k\geq0}\Bigl(v_{k+1}\frac{\p}{\p v_k}+r_{k+1}\frac{\p}{\p r_k}\Bigr).
\eeq
By using the Virasoro constraints~\eqref{Lminus1}--\eqref{openVirasoro} we will show in Section~\ref{sec3}
the following theorem.
\begin{thm}\label{main1}
The series $\Delta F=\sum_{g\geq1}\e^{2g-2}F^c_g+\sum_{p\geq1}\e^{p-1}F^o_p$ is the unique solution to the following equation:
\begin{align}
&\sum_{k\geq0}\frac{\p\Delta F}{\p v_k}\bigg(\p^k(A) +\sum_{j=1}^k\binom k j\p^{j-1}\big(A^{\frac12}\big) \p^{k+1-j} \big(A^{\frac12}\big)\bigg)
\nn \\
&
+\frac12\sum_{k\geq0}\frac{\p\Delta F}{\p r_k }\bigg(\p^k (r A B)+\sum_{j=1}^k\binom k j\p^{j-1}\big(A^{\frac12}\big) \p^{k+1-j}\big(r A^{\frac12} B \big)\bigg)
\nn \\
=&
\frac{\e^2}2\sum_{k,\ell\geq0}\bigg(\frac{\p^2\Delta F}{\p v_k \p v_\ell}+\frac{\p\Delta F}{\p v_k}\frac{\p\Delta F}{\p v_\ell}\bigg)\p^{k+1}\big(A^{\frac12}\big)\p^{\ell+1}\big(A^{\frac12}\big)
\nn \\
&+\frac{\e^2}2\sum_{k,\ell\geq0}\left(\frac{\p^2\Delta F}{\p v_k \p r_\ell}+\frac{\p\Delta F}{\p v_k}\frac{\p\Delta F}{\p r_\ell}\right)\p^{k+1}\big(A^{\frac12}\big)\p^{\ell+1} \big(r A^{\frac12} B \big)
\nn\\
&+\frac{\e^2}8\sum_{k,\ell\geq0}\left(\frac{\p^2\Delta F}{\p r_k \p r_\ell}+\frac{\p\Delta F}{\p r_k}\frac{\p\Delta F}{\p r_\ell}\right)\p^{k+1}\big(r A^{\frac12} B \big) \p^{\ell+1} \big(r A^{\frac12} B \big) 
\nn\\
&+\sum_{k\geq0}\frac{\p\Delta F}{\p v_k}\Big(\frac{\e^2}{16}\p^{k+2}\big(A^2\big)+\e r A^{\frac12} B \p^{k+1}\big(A^{\frac12}\big)\Big)
\nn \\
&+\sum_{k\geq0}\frac{\p\Delta F}{\p r_k}
\bigg(
\e^2\p^{k+1}\Big(\frac{A}{16r}\p(r^2 AB)+\frac{r}{12}\p\big(B^3\big)\Big) 
+\e\Big(\frac14 r A^{\frac12} B   \p^{k+1}\big(r A^{\frac12}B\big)+\frac34\p^{k+1}\big(B^2\big)\Big)
\bigg)
\nn\\
&+\e
\Big(
\frac{A}{16r}\p(r^2 AB)+\frac{r}{12}\p\big(B^3\big)
\Big)
+\frac{A^2}{16}+B^2-\frac{1}{4} AB.
\label{loopequation}
\end{align}
\end{thm}

\begin{emp}\label{emp1}
Denote $u:=v+\frac{r^2}2$ and $u_k:=\p^k(u)$ for $k\geq0$.
From Theorem~\ref{main1} and the formulae \eqref{Fc12},
we obtain the following expressions:
\begin{align*}
F^o_1=&\frac12\log \left(v_1+rr_1\right)=\frac12\log u_1, \\
F^o_2=&
\frac1{24 v_1 \left(v_1+r r_1\right)^3}
\Big(
-r^2 v_2 r_1^3+3 r^2 r_3 v_1^2-4 r^3 v_1 r_2^2+4 v_1^3 r_2-3 r v_1 r_1^4+v_1^2 r_1^3+3 r^3 r_3 v_1 r_1
\\
&+6 r^2 v_1 r_1^2 r_2+18 r v_1^2 r_1 r_2+3 v_3 r^2 v_1 r_1+3 v_3 r v_1^2-8 r^2 v_1 v_2 r_2-9 r v_1 v_2 r_1^2-4 r v_1 v_2^2
\Big)
\\
=&
-\frac{r u_2^2}{6u_1^3}+\frac{r_1 u_2+3r u_3}{24 u_1^2}+\frac{r_1(r_1^2-u_2)}{24 (u_1-rr_1) u_1}+\frac{r_2}{8u_1}-\frac{r_2}{24 (u_1-rr_1)}.
\end{align*}
\end{emp}

Analogous to the work of Itzykson--Zuber \cite{IZ},
we introduce the following variables:
\begin{align}
&I_{k}:=\sum_{n\geq0}t_{n+k}\frac{v_{\rm top}^n}{n!}, \label{defI1}\\
&J_{k}:=\sum_{n\geq0}t_{n+k+1}\sum_{i=0}^n \frac{v_{\rm top}^i r_{\rm top}^{2n-2i+1}}{i!(2n-2i+1)!!}
+\sum_{n\geq0}\frac{s_{n+k}}{n!}\Big(v_{\rm top}+\frac{r_{\rm top}^2}{2}\Big)^n. \label{defI2}
\end{align}
We will show in Section \ref{sec2} that the transformation from ${\bf t}, {\bf s}$ to $I_k,J_k,k\geq0$ is invertible.
By using Theorem \ref{main0}, we will prove in Section \ref{sec4} that
for $p\geq2$, the genus $p$ open free energy $\mathcal F^o_p$ can be expressed as 
an element in
$\mathbb{C}[(1-I_1)^{-1}, (1-I_1-J_0J_1)^{-1}, J_0, J_1, I_2,J_2,\dots,I_{2p-1},J_{2p-1}]$. 
As an application of these expressions, we will prove in Section \ref{sec4} the following corollary.
\begin{cor}\label{cor1}
Let $m,n\geq0$, $k_1,\dots,k_m\geq2,\ell_1,\dots,\ell_n\geq2$ and $d\geq0$, $p\ge2$.
If $m+n+d>13p-7$,
the following identities hold true:
\beq\label{exprFostarp}
\sum_{q\geq0}\frac{(-1)^q}{(d-q)!q!(7p-4-q)!}
\left.
\Lambda_{k_1,\dots,k_m;1^{d-q},\ell_1,\dots,\ell_n}\big(\mathcal F^o_p\big)
\right|_{s_0=1,t_{\geq0}=s_{\geq1}=0}
=0.
\eeq
Here
$\Lambda_{k_1,\dots,k_m;\ell_1,\dots,\ell_n}$ are operators defined as follows: 
\begin{align}
&\Lambda_{k_1,\dots,k_m;\ell_1,\dots,\ell_n}:=
\prod_{i=1}^m \bigg(\frac{\p}{\p t_{k_i}}-\sum_{j=0}^{k_i-1}\frac{(-1)^j}{2^j (2j+1) j!} \frac{\p}{\p s_{k_i-j-1}}\bigg)
\prod_{i=1}^n \bigg(\sum_{j=0}^{\ell_i}\frac{(-1)^j}{2^j j!}\frac{\p}{\p s_{\ell_i-j}}\bigg).
\label{defLambda}
\end{align}
\end{cor}

We also have the following conjectural statement: 
for $m,n\geq0$, $k_1,\dots,k_m\geq2,\ell_1,\dots,\ell_n\geq2$ and $d\geq0$, $p\ge2$, 
the identities
\beq\label{conjecturalLambda}
\sum_{q\geq0}\frac{(-1)^q}{(d-q)!q!(3p-3-q)!}
\Lambda_{k_1,\dots,k_m;1^{d-q},\ell_1,\dots,\ell_n}\big(\mathcal F_p^o\big)
\mid_{s_0=1,t_{\geq0}=s_{\geq1}=0}
=0
\eeq
hold true when $m+n+d>6p-6$.
This is stronger than \eqref{exprFostarp}.

\medskip

\noindent 
{\bf Organization of the paper.} In Section~2, we prove Theorem~\ref{main0}.
In Section~3, we prove Theorem~\ref{main1}.
In Section~4, we present some applications.

\smallskip

\noindent {\bf Acknowledgements.}
We are grateful to Professor Youjin Zhang his advice and for suggesting this project. We also thank Professor Si-Qi Liu for helpful discussions. The work is partially supported by NSFC grants (No. 12371254, No. 12322111), by National Key R and D Program of China 2020YFA0713100, and by CAS No. YSBR-032.

\section{Jet-variable representation}\label{sec2}

In this section, we will prove Theorem~\ref{main0}.

Let $\Omega_{m,n}$, $m,n\ge0$, be the 
two-point correlation functions of the KdV hierarchy (cf.~\cite{BDY, DZ-norm}).
Recall that, for an arbitrary solution $(w,\rho)$ to the Burgers-KdV hierarchy, 
the defining equations of the associated tau-tuple $(\tau_1,\tau_2)$ \cite{YZ} (cf.~\cite{BDY, Bu2, DY, DZ-norm})  
are given by 
\beq\label{deftau}
\e^2\frac{\p^2\log \tau_1}{\p t_m\p t_n}=\Omega_{m,n},\quad
\e\frac{\p\log \tau_2}{\p t_n}=R_n,\quad
\e\frac{\p\log \tau_2}{\p s_n}=Q_n,\quad m,n\geq0.
\eeq
We denote
\begin{align}
&\langle\langle \tau_{k_1}\cdots\tau_{k_m}\rangle\rangle^c
:=\e \frac{\p^{m} \log \tau_1}{\p t_{k_1}\cdots \p t_{k_m}},\label{correlationfunction1} \\
&\langle\langle \tau_{k_1}\cdots\tau_{k_m}\sigma_{\ell_1}\cdots \sigma_{\ell_n}\rangle\rangle^o
:=\e \frac{\p^{m+n} \log \tau_2}{\p t_{k_1}\cdots \p t_{k_m} \p s_{\ell_1} \cdots \p s_{\ell_n}}. \label{correlationfunction2}
\end{align}
It is well known that $\langle\langle \tau_{k_1}\cdots\tau_{k_m}\rangle\rangle^c$ are differential polynomials of $w$.
By the definition \eqref{deftau} and the hierarchy \eqref{openkdv1}--\eqref{openkdv2},
we know that $\langle\langle \tau_{k_1}\cdots\tau_{k_m}\sigma_{\ell_1}\cdots \sigma_{\ell_n}\rangle\rangle^o$ 
are also differential polynomials in $\mathcal A[\e]$.

It follows from equation~\eqref{string} that 
the topological solution $(w_{\rm top},\rho_{\rm top})$ satisfies the following Euler--Lagrange equations:
\begin{align}
&\sum_{n\geq0}\tilde t_{n+1} K_n(\underline{w};\e)+t_0=0, \label{elkdv}\\
&\sum_{n\geq0}\bigl(\tilde t_{n+1} R_n(\underline{w},\underline{\rho};\e)+s_{n+1}Q_n(\underline{w},\underline{\rho};\e)\bigr)+s_0=0.
\label{el}
\end{align} 
Here, 
$\underline{w}=(w,w_x,w_{xx},\dots)$ and $\underline{\rho}=(\rho,\rho_x,\rho_{xx},\dots)$, and 
for simplifying the notations we use $(w,\rho)$ to denote $(w_{\rm top},\rho_{\rm top})$.

By taking $\e=0$ in the system of equations~\eqref{elkdv} and \eqref{el}, 
we have that the pair
\[
(v,r):=(w,\rho)|_{\e=0}
\]
satisfies the following genus zero Euler--Lagrange equations:
\begin{align}
&\sum_{n\geq0} \tilde t_{n}  \frac{v^n}{n!}=0, \label{g0elkdv}\\
&\sum_{n\geq0} \tilde t_{n+1} \sum_{i=0}^n \frac{v^i r^{2n-2i+1}}{i!(2n-2i+1)!!}
+\sum_{n\geq0} \frac{s_n}{n!} \biggl(v+\frac{r^2}2\biggr)^n=0. \label{g0el}
\end{align}
Then it is easy to see from the definitions \eqref{defI1}, \eqref{defI2} that 
\begin{align}
&I_0=v,\quad J_0=r, \label{Ijet1}\\
&I_k=\frac{\p_x (I_{k-1})-\delta_{k,1}}{v_1}, 
\quad
J_k=\frac{\p_x (J_{k-1})-r_1 I_k}{v_1+r r_1},
\label{Ijet2}
\end{align}
where $k\geq1$.
Straightforwardly, we have the following lemma.
\begin{lem}\label{lemIJform}
The functions $I_k, J_k, k\ge 1 $, have the form
\begin{align}
&I_k=\sum_{\substack{\lambda\in \mathcal{P}_{k-1}}}
c'_{k; \lambda} v_1^{-k-\ell(\lambda)} v_{\lambda+1},
\\
&J_k=\sum_{1\le d \le k}\sum_{\substack{ \lambda, \mu\in\mathcal{P} \\ 
|\lambda|+|\mu|=k-d}} 
c''_{k;\lambda;\mu}(v_1^{-1},u_1^{-1}) r_d v_{\lambda+1} u_{\mu+1},
\end{align}
where $c'_{k; \lambda}$ are rational numbers, and 
$c''_{k;\lambda;\mu}(z_1,w_1)$ are homogeneous polynomials of $z_1,w_1$ with degree $k+\ell(\lambda)+\ell(\mu)+1$
and with rational coefficients. Here we denote by $\mathcal P$ the set of partitions 
\[
\mathcal P=\{(\lambda_1,\lambda_2,\dots)| \lambda_1 \ge \lambda_2 \ge \cdots \ge 1, \sum_i \lambda_i<\infty \},
\]
by $\ell(\lambda)$ the length of $\lambda$, 
by $|\lambda|$ the summation of the components of $\lambda$,
and by $\mathcal P_k$ the set of partitions $\lambda$ with $|\lambda|=k$.
We also denote $a_\lambda:=a_{\lambda_1}\cdots a_{\lambda_{\ell(\lambda)}}$.
\end{lem}

For example,
\begin{align*}
I_1=1-\frac1{v_1},\quad 
I_2=\frac{v_2}{v_1^3}, \quad 
J_1=\frac{r_1}{v_1 u_1},\quad 
J_2=\frac{r_2}{v_1 u_1^2}-r_1\Bigl(\frac{u_2}{v_1 u_1^3}+\frac{v_2}{v_1^3 u_1}+\frac{v_2}{v_1^2 u_1^2}\Bigr),
\end{align*}
and conversely,
\begin{align*}
v_1=&\frac{1}{U_1},\quad
r_1=\frac{J_1}{U_1U_2},\quad
v_2=\frac{I_2}{U_1^3},\quad
r_2=&\frac{J_2}{U_2^3}+\frac{J_1^3}{U_1^2 U_2^3}+J_1 I_2\Bigl(\frac{1}{U_1 U_2^3}+\frac1{U_1^2 U_2^2}+\frac1{U_1^3 U_2}\Bigr),
\end{align*}
where we denote
\beq\label{defU1U2}
U_1:=1-I_1, \quad U_2:=1-I_1-J_0J_1.
\eeq
The following lemma gives the inverse of \eqref{defI1}--\eqref{defI2}, which generalizes the one in~\cite{DY, Zhou}.
\begin{lem}\label{lemst}
For~$k\geq0$, we have
\begin{align}
&t_k=\sum_{n\geq0}\frac{(-1)^n}{n!}I_0^n I_{n+k},\label{ti} \\
&s_k=\sum_{n\geq0}\sum_{i=0}^n \frac{(-1)^{n+1}I_0^i J_0^{2n-2i+1}}{2^{n-i}i!(n-i)!(2n-2i+1)}I_{n+k+1}
+\sum_{n\geq0}\frac{(-1)^n}{n!}\Bigl(I_0+\frac12 J_0^2\Bigr)^n J_{n+k}. \label{sij}
\end{align}
\end{lem}
\begin{prf}
See in~\cite{Zhou} the proof for~\eqref{ti}.
To prove \eqref{sij}, let us denote
\beq
\alpha_n:=\sum_{i=0}^n \frac{(-1)^{n+1}I_0^i J_0^{2n-2i+1}}{2^{n-i}i!(n-i)!(2n-2i+1)},
\quad
\beta_n:=\frac{(-1)^n}{n!}\Bigl(I_0+\frac12 J_0^2\Bigr)^n.
\eeq
Substitute \eqref{ti} and \eqref{sij} into~\eqref{defI2} and then we need to verify
\begin{align}
J_k=&\sum_{n,m\geq0}\frac{(-1)^m I_0^m}{m!} I_{n+m+k+1}\sum_{i=0}^n \frac{I_0^i J_0^{2n-2i+1}}{i!(2n-2i+1)!!} \nn \\
&\quad+\sum_{n,m\geq0}\frac{1}{n!}\Big(I_{0}+\frac{J_0^2}{2}\Big)^n\left(\alpha_m I_{n+m+k+1}+\beta_m J_{n+m+k}\right),\quad k\geq0 . \label{identityIZ}
\end{align}
Equivalently, it suffices to check the following identities:
\begin{align}
&\sum_{n=0}^k \frac{1}{n!}\Big(I_0+\frac12 J_0^2\Big)^n \alpha_{k-n}
=-\frac{J_0^{2k+1}}{(2k+1)!!}
=-\sum_{n=0}^k \frac{(-1)^{k-n}I_0^{k-n}}{(k-n)!}\sum_{i=0}^n \frac{I_0^i J_0^{2n-2i+1}}{i!(2n-2i+1)!!}, \\
&\sum_{n=0}^k\frac{1}{n!}\Big(I_0+\frac12 J_0^2\Big)^n \beta_{k-n}=\delta_{k,0}.
\end{align}
The lemma is proved.
\end{prf}

It follows from Lemma \ref{lemst} and the relations \eqref{Ijet1}--\eqref{Ijet2} that
\begin{align}
t_k=&v_k-\delta_{k,1}+\text{higer-order terms in} \, \mathbb C[[v, v_1-1, v_2, \dots]], \\
s_k=&r_k+\text{higer-order terms in} \, \mathbb C[[v, r, v_1-1, r_1, v_2, r_2, \dots]].
\end{align}
By applying this transformation, the free energies $\mathcal F^o_p$ can be expressed as an element 
\beq\label{FpRing}
F_p^o  \in \CC[[v, r, v_1-1,  r_1, v_2, r_2, \dots]].
\eeq

\begin{lem}\label{decomposelem}
For $n\geq0$, the differential polynomials $R_n$ and $Q_n$ can be expanded into the form
\begin{align}
&R_n=\sum_{i=0}^n \bigg(a_i(\underline{w},\underline{\rho};\e) \sum_{j=0}^{n-i}\frac{w^j \rho^{2n-2i-2j+1}}{j!(2n-2i-2j+1)!}
+b_i(\underline{w},\underline{\rho};\e) \frac{w^{n-i+1}}{(n-i+1)!}\bigg), \label{decomR}\\
&Q_n=\sum_{i=0}^{n+1} a_i(\underline{w},\underline{\rho};\e) \frac{(w+\frac12\rho^2)^{n-i+1}}{(n-i+1)!},  \label{decomQ}
\end{align}
where the coefficients $a_i$ and~$b_i$ are certain elements in $\mathcal A[\e]$, which do not explicitly contain~$w$.
Moreover, $a_i$'s and~$b_i$'s have the following form:
\beq\label{abform}
a_i(\underline{w},\underline{\rho};\e)=\sum_{j=[\frac{i+1}2]}^{2i-1}\e^j a_{i,j}(\underline{w},\underline{\rho}),
\quad
b_i(\underline{w},\underline{\rho};\e)=\sum_{j=[\frac{i+1}2]}^{2i-2}\e^j b_{i,j}(\underline{w},\underline{\rho}).
\eeq
\end{lem}
\begin{prf}
By using formula~\eqref{defK}, one can obtain
\beq
K_n(\underline{w};\e)=\sum_{i\geq0}c_i(\underline{w};\e) \frac{w^{n-i+1}}{(n-i+1)!}.
\eeq
where the coefficients $c_i=c_i(\underline{w};\e)$ satisfy
\begin{align}
\Big(i-\frac12\Big)\p_x c_i=&\bigg(\Big(-i+\frac12\Big)w_x+\frac18 \e^2\p_x^3\bigg)c_{i-1}
+\frac18\e^2\Big(w_{xxx}+3w_{xx}\p_x+3w_x\p_x^2\Big)c_{i-2}  \nn \\
&+\frac38\e^2\left(w_xw_{xx}+w_x^2\p_x\right)c_{i-3}+\frac18\e^2 w_x^3 c_{i-4}.
\end{align}
By substituting \eqref{decomR}--\eqref{decomQ} into \eqref{defR}--\eqref{defQ} respectively, we have
\begin{align}
ia_i=&\Big(\frac12\e\rho_x+\e\rho\p_x+\frac12\e^2\p_x^2\Big)a_{i-1} 
+\Big(\e\rho u_x+\frac12(w_x+\rho\rho_x)_{x}+\e^2 (w_x+\rho\rho_x) \p_x\Big)a_{i-2} \nn \\
&+\frac12\e^2 (w_x+\rho\rho_x)^2 a_{i-3}, \\
\Big(i-\frac12\Big) b_i
=&\Big(\frac12\rho^2+\frac12\e\rho_x+\frac12\e^2\p_x^2\Big)b_{i-1}
+\Big(\e\rho w_x+\frac12 w_{xx}+\e^2 w_x\p_x\Big) b_{i-2} \nn\\
&+\frac12\e^2 w_x^2 b_{i-3}
-\frac12\rho a_{i-1}
+\Big(\e\rho\rho_x+\frac12\e^2 \rho_{xx}+\e^2\rho_x\p_x\Big)a_{i-2} \nn \\
&+\frac12\e^2\rho_x (2w_x+\rho\rho_x) a_{i-3}
+\Big(\frac12\rho+\frac34\e\p_x\Big)c_{i-1}
+\frac34 \e w_x c_{i-2}.
\end{align}
The lemma can then be proved by induction.
\end{prf}
\begin{emp}\label{empab}
The first several coefficients $a_i$ and $b_i$ are as follows:
\begin{align*}
&a_0=1,\quad a_1=\frac12\e\rho_x,\\
&a_2=\frac12 \e\left(\rho w_x+\rho^2\rho_x\right)+\frac18\e^2\left(3\rho_x^2+2w_{xx}+4\rho\rho_{xx}\right)+\frac18\e^3 \rho_{xxx}. \\
&b_0=0,\quad b_1=\frac12\e\left(w_x+\rho\rho_x\right)+\frac13\e^2\rho_{xx}, \\
&b_2=\frac1{24}\e^2\left(12w_x\rho_x+15\rho\rho_x^2+4\rho w_{xx}+4\rho^2 \rho_{xx}\right)
+\frac1{24}\e^3\left(16\rho_x\rho_{xx}+3w_{xxx}+5 \rho\rho_{xxx}\right)+\frac1{15}\e^4 \rho^{(4)}.
\end{align*}
\end{emp}
\begin{lem}\label{lemv}
Let 
$\Delta \underline{v}=(\Delta v_0,\Delta v_1,\dots)$
and
$\Delta \underline{r}=(\Delta r_0,\Delta r_1,\dots)$.
Then we have
\begin{align}
&\sum_{n\geq 0}\left(t_{n+1}R_n(\underline{v}({\bf t})+\Delta \underline{v},\underline{r}({\bf t},{\bf s})+\Delta\underline{r};\e)
+s_{n+1}Q_n(\underline{v}({\bf t})+\Delta\underline{v},\underline{r}({\bf t},{\bf s})+\Delta\underline{r};\e)\right)+s_0 \nn \\
=&\sum_{m,n\geq0} \sum_{\substack{k_1,\dots,k_m\geq0\\ \ell_1,\dots,\ell_n\geq0}}
\mathcal{V}_{k_1,\dots,k_m;\ell_1,\dots,\ell_n}\left({\bf t},{\bf s};\e\right)
\frac{\Delta v_{k_1}\cdots \Delta v_{k_m}\Delta r_{\ell_1}\cdots \Delta r_{\ell_n}}{m!n!},
\label{defV}
\end{align}
where the coefficients $\mathcal{V}_{k_1,\dots,k_m;\ell_1,\dots,\ell_n}$ have the following expressions:
\[
\mathcal{V}_{k_1,\dots,k_m;\ell_1,\dots,\ell_n}({\bf t},{\bf s};\e)=\sum_{p\geq0}\e^p V^{[p]}_{k_1,\dots,k_m;\ell_1,\dots,\ell_n}(\underline{v}({\bf t}),\underline{r}({\bf t},{\bf s}))
\]
with
\beq \label{VVbelong}
V^{[p]}_{k_1,\dots,k_m;\ell_1,\dots,\ell_n}(\underline{v}, \underline{r} )\in
\sum_{\substack{i,j\geq0\\i+j\leq 4p+2c_0-1}}
v_1^{-i} (v_1+r r_1)_1^{-j}\CC[r,v_1,r_1,\dots,v_{2p+c_0},r_{2p+c_0}].
\eeq
Here $c_0$ counts the number of $0$ in $(k_1,\dots,k_m,\ell_1,\dots,\ell_n)$.
\end{lem}
\begin{prf}
By using Lemma~\ref{decomposelem}, we can express 
$\mathcal{V}_{k_1,\dots,k_m;\ell_1,\dots,\ell_n}({\bf t},{\bf s};\e)$
in terms of $a_{k,i},b_{k,i}$ and $I_k,J_k$ as follows:
\begin{align}
\mathcal V_{\varnothing;\varnothing}
=&
\sum_{n\geq0}
\Bigg(
t_n \sum_{k\geq 0}a_k(\underline{v},\underline{r};\e) \sum_{j=0}^{n-k-1}\frac{v^j r^{2n-2k-2j-1}}{j!(2n-2k-2j-1)!!}
+t_n\sum_{k\geq 0}b_k(\underline{v},\underline{r};\e) \frac{v^{n-k-1}}{(n-k-1)!}\nn \\
&+s_n\sum_{k\geq0}a_k(\underline{v},\underline{r};\e) \frac{(v+\frac12 r^2)^{n-k}}{(n-k)!}
\Bigg)
\nn\\
=&
\sum_{k\geq0}\sum_{j=[\frac{k+1}{2}]}^{2k-1} \e^j a_{k,j}(\underline{v},\underline{r}) J_{k}+\sum_{k\geq 0} \sum_{j=[\frac{k+1}{2}]}^{2k-2} \e^j b_{k,j}(\underline{v},\underline{r}) I_{k+1},
\label{Ag1}
\\
\mathcal V_{0;\varnothing}
=&\sum_{k\geq0} \Bigg(\sum_{j=[\frac{k+1}{2}]}^{2k-1} \e^j a_{k,j}(\underline{v},\underline{r})J_{k+1}+\sum_{j=[\frac{k+1}{2}]}^{2k-2} \e^j b_{k,j}(\underline{v},\underline{r}) I_{k+2}\Bigg),
\\
\mathcal V_{\varnothing;0}
=&\sum_{k\geq0}\Bigg(\sum_{j=[\frac{k+1}{2}]}^{2k-1} \e^j \frac{\p a_{k,j}(\underline{v},\underline{r})}{\p r} J_{k}
+r \sum_{j=[\frac{k+1}{2}]}^{2k-1} \e^j a_{k,j}(\underline{v},\underline{r}) J_{k+1} \nn \\
&+\bigg(\sum_{j=[\frac{k+1}{2}]}^{2k-1} \e^j a_{k,j}(\underline{v},\underline{r}) 
+\sum_{j=[\frac{k+1}{2}]}^{2k-2} \e^j \frac{\p b_{k,j}(\underline{v},\underline{r})}{\p r} \bigg)I_{k+1}\Bigg),
\\
&\cdots \quad \cdots
\nn
\end{align}
Then by using Lemma \ref{lemIJform}, the lemma is proved.
\end{prf}

~\\

Now we are ready to prove Theorem~\ref{main0}. 

\begin{prfn}{Theorem \ref{main0}}
Using Lemma \ref{lemv} 
and taking
\beq
\Delta v_k=\sum_{g\geq1}\e^{2g}\p_x^{k+2}\mathcal F^c_{g},\quad \Delta r_k=\sum_{p\geq1}\e^p \p_x^{k+1}\mathcal F^o_p,
\eeq
we have
\begin{align}
\sum_{m,n\geq0}\sum_{\substack{k_1,\dots,k_m\geq0,\\ \ell_1,\dots,\ell_n\geq0}}
\frac{\mathcal{V}_{k_1,\dots,k_m;\ell_1,\dots,\ell_n}}{m!n!} 
\Delta v_{k_1}\cdots \Delta v_{k_m}
\Delta r_{\ell_1}\cdots \Delta r_{\ell_n}-r_0-\Delta r_0=0.
\label{el1}
\end{align}

Then by taking the coefficient of $\e^p$ of \eqref{el1}, we have the following recursion:
\begin{align}
-\frac{1}{v_1+rr_1}\p_x  F^o_p
=\widetilde \sum
\frac{V_{k_1,\dots,k_m;\ell_1,\dots,\ell_n}^{[q]}}{m!n!}
\p_x^{k_1+2} F^c_{g_1}\cdots \p_x^{k_m+2}  F^c_{g_m}
\p_x^{\ell_1+1} F^o_{p_1}\cdots \p_x^{\ell_n+1} F^o_{p_n},\label{el3}
\end{align}
where the summation $\widetilde\sum$ is over
\begin{align*}
&k_1,\dots,k_m,\ell_1,\dots,\ell_n\geq0,\quad g_1,\dots,g_m,p_1,\dots,p_n\geq1,
\\
&
k_1+\cdots+k_m+\ell_1 + \cdots + \ell_n\leq q,
\quad
q+2g_1+\cdots+2g_m+p_1+\cdots+p_n=p,
\end{align*}
and we used the result
\begin{align}
V_{\varnothing;0}^{[0]}=I_{1}+J_0 J_1=1+\frac{1}{v_1+rr_1}.
\end{align}
By using formulae \eqref{VVbelong}, \eqref{el3} and the fact that
\beq
\p_x^2 F^c_p\in v_1^{-(4p-4)}\CC[v_1,v_{2},\dots,v_{3p}], \quad p\geq1,
\eeq
we can obtain recursively from~\eqref{el3} that
\beq\label{pfog}
\p_x F^o_p\in 
v_1^{-(5p-1)}(v_1+rr_1)^{-(5p-1)}\CC[r,v_1,r_1,\dots,v_{2p},r_{2p}].
\eeq

It follows from \eqref{FpRing} and \eqref{pfog} that
\beq\label{FoRing}
F^o_p
\in \CC[\log v_1, \log (v_1+rr_1), v_1^{-1}, (v_1+rr_1)^{-1}, r, v_1, r_1, v_2,r_2,\dots,v_{2p-1},r_{2p-1}].
\eeq
For $p=1$, formula~\eqref{el3} reads
\beq
\p_x F^o_1=-(v_1+rr_1)V_{\varnothing;\varnothing}^{[1]}
=\frac12\p_x\log(v_1+rr_1).
\eeq
Hence \eqref{Fo1} is proved. 

Now let us consider $p\geq2$. 
In this case, we will use two scaling invariant properties.
From the equations \eqref{VirasoroKdV} and \eqref{Virasoroconstraints}, we obtain
\beq\label{ScalingSymmetry}
\sum_{n\geq0}\frac{2n+1}{2}\tilde t_n \frac{\p \mathcal F^o_p}{\p t_n}+\sum_{n\geq0}(n+1)s_n\frac{\p \mathcal F^o_p}{\p s_n}=0.
\eeq
From \eqref{openkdv1}--\eqref{defQ},
we find that when $m+n\geq1$,
$\langle\langle \tau_{k_1}\cdots\tau_{k_m}\sigma_{\ell_1}\cdots \sigma_{\ell_n}\rangle\rangle^o \in \mathcal A[\e]$
satisfy
\begin{align}
&\sum_{k\geq0} \left(k w_{kx} \frac{\p}{\p w_{kx}}+k \rho_{kx} \frac{\p}{\p \rho_{kx}}-\e\frac{\p}{\p \e}\right)
\langle\langle \tau_{k_1}\cdots\tau_{k_m}\sigma_{\ell_1}\cdots \sigma_{\ell_n}\rangle\rangle^o 
\nn \\
=&(m+n-1)\langle\langle \tau_{k_1}\cdots\tau_{k_m}\sigma_{\ell_1}\cdots \sigma_{\ell_n}\rangle\rangle^o,
\\
&\sum_{k\geq0} \left((k-1) w_{kx} \frac{\p}{\p w_{kx}}+\big(k-\frac12\big) \rho_{kx} \frac{\p}{\p \rho_{kx}}-\frac32\e\frac{\p}{\p \e}\right)
\langle\langle \tau_{k_1}\cdots\tau_{k_m}\sigma_{\ell_1}\cdots \sigma_{\ell_n}\rangle\rangle^o 
\nn\\
=&-\big(k_1+\cdots+k_m+\ell_1+\cdots+\ell_n-m-\frac n2+\frac32
\big)
\langle\langle \tau_{k_1}\cdots\tau_{k_m}\sigma_{\ell_1}\cdots \sigma_{\ell_n}\rangle\rangle^o.
\end{align}
Note that
\beq
\left. \p_x^k w_{\rm top} \right|_{{\bf t}=0}=\delta_{k,1},
\quad
\left. \p_x^k \rho_{\rm top} \right|_{{\bf t}={\bf s}=0}=0.
\eeq
Then by  taking ${\bf t}={\bf s}=0$ in \eqref{correlationfunction2} and comparing the coefficients of each power of $\e$ in the both sides,
we obtain that
\beq
\langle \tau_{k_1}\cdots\tau_{k_m}\sigma_{\ell_1}\cdots \sigma_{\ell_n}\rangle^o_p\neq0,
\eeq
only if
\beq
k_1+\cdots+k_m+\ell_1+\cdots+\ell_n-m-\frac n2=\frac32(p-1).
\eeq
Equivalently, we have
\beq\label{tautopcond}
\sum_{k\geq0} \bigg( (k-1)t_k\frac{\p \mathcal F_p^o}{\p t_k}
+\big(k-\frac12\big)s_k\frac{\p \mathcal F_p^o}{\p s_k}\bigg)
=\frac32(p-1)\mathcal F^o_p.
\eeq
Together with \eqref{ScalingSymmetry}, we arrive at
\beq\label{dilaton}
\sum_{k\geq1}\bigg( \tilde t_k \frac{\p \mathcal F_p^o}{\p t_k}+s_k \frac{\p \mathcal F_p^o}{\p s_k} \bigg)=-(p-1) \mathcal F^o_p.
\eeq
The conditions \eqref{homogeneous1} and \eqref{homogeneous2} follow from \eqref{dilaton} and \eqref{tautopcond} respectively. 
Then \eqref{Fog} follows from \eqref{homogeneous1} and \eqref{FoRing}.
The theorem is proved.
\end{prfn}

We also have the conjectural form for $F_p^o$ with $p\geq2$:
\begin{align}
F^o_p=
\frac{v_1^{4p-4}}{(v_1+rr_1)^{3p-3}}\sum_{i=1}^{3p-3} 
\sum_{\substack{\lambda, \mu\in\mathcal{P}\\ \lambda_i\leq 2p-2\\ \mu_i \leq 2p-1}}
\sum_{\substack{|\lambda|+|\mu|= \\\frac32(p-1)+\frac n 2+\frac i 2}}
C^{[i]}_{\lambda;\mu}r^i \frac{v_{\lambda+1}r_{\mu}}{v_1^{|\lambda|+\ell(\lambda)+|\mu|}},
\label{cnjFo}
\end{align}
where $C^{[i]}_{\lambda;\mu}$ are certain rational numbers.

\section{Loop equation}\label{sec3}

In this section, 
we derive the Dubrovin--Zhang type loop equation for the topological tau-function 
of the Burgers--KdV hierarchy using the Virasoro constraints.

Consider the generating series
\beq\label{generateL}
\mathcal L(\lambda):=\sum_{m\geq-1}\frac{\mathcal L_m^{\rm ext}}{\lambda^{m+2}},
\eeq
where $\L_m^{\rm ext}$ are the operators given by~\eqref{openVirasoro}.
It is easy to verify that 
the series $\L(\lambda)$ can be equivalently written as
\beq\label{lij}
\mathcal L(\lambda)=\left[\I_1(\lambda)\J_1(\lambda)+\I_2(\lambda)\J_2(\lambda)\right]_{-}
+\frac{\e^2}{2}\J_1(\lambda)^2
-\frac{3\e}{4}\p_\lambda \J_2(\lambda)
+\frac{1}{\lambda}\left(\frac{t_0^2}{2\e^2}+\frac{s_0}{\e}\right)+\frac{5}{8\lambda^2},
\eeq
where $[\,\,]_{-}$ means taking the terms with negative powers of~$\lambda$, and
\begin{align}
&\I_1(\lambda):=\sum_{n\geq0}\frac{2^n\lambda^{n-\frac12}}{(2n-1)!!}\tilde t_n, 
\quad \I_2(\lambda):=\sum_{n\geq0}\frac{\lambda^n}{n!} s_n,\\
&\J_1(\lambda):=\sum_{n\geq0}\frac{(2n+1)!!}{2^{n+1}\lambda^{n+\frac32}}
\frac{\p}{\p t_n},\quad
\J_2(\lambda):=\sum_{n\geq0}\frac{(n+1)!}{\lambda^{n+2}}\frac{\p}{\p s_n}.
\label{defJ}
\end{align}
We have the following lemma.

\begin{lem}
The constraints~\eqref{Virasoroconstraints} are equivalent to the following equations:
\begin{align}
&\left[\left(\I_1(\lambda)\J_1(\lambda)+\I_2(\lambda)\J_2(\lambda)\right)(\F^{\rm c}_0)\right]_{-}
+\frac12 \left(\J_1(\lambda)(\F^{\rm c}_0)\right)^2+\frac{t_0^2}{2\lambda}=0, \label{JF1}\\
&\left[\left(\I_1(\lambda)\J_1(\lambda)+\I_2(\lambda)\J_2(\lambda)\right)(\F^{\rm o}_0)\right]_{-}
+ \J_1(\lambda)(\F^{\rm c}_0)\J_1(\lambda)(\F^{\rm o}_0)+\frac{s_0}{\lambda}=0, \label{JF2}\\
&\left[\left(\I_1(\lambda)\J_1(\lambda)+\I_2(\lambda)\J_2(\lambda)\right)(\Delta\F)\right]_{-}
+\J_1(\lambda)(\F^{\rm c}_0)\J_1(\lambda)(\Delta\F)\nn\\
&\quad +\frac{\e^2}2\left(\J_1(\lambda)^2(\Delta \F)+\left(\J_1(\lambda)(\Delta\F)\right)^2\right)
+\e\left(\J_1(\lambda)(\F^{\rm o}_0)\J_1(\lambda)(\Delta\F)-\frac34\p_\lambda \J_2(\lambda)(\Delta\F)\right)\nn\\
&\quad +\frac12\J_1(\lambda)^2(\F^{\rm c}_0)+\frac{\e}{2}\J_1(\lambda)^2(\F^{\rm o}_0)
+\frac12\J_1(\lambda)(\F^{\rm o}_0)^2-\frac34\p_\lambda \J_2(\lambda)(\F^{\rm o}_0)+\frac{5}{8\lambda^2}=0,
\label{loop0}
\end{align}
where $\Delta\F$ is defined as
\beq\label{deltaF}
\Delta\F:=\sum_{g\geq1}\e^{2g-2}\F^c_g+\sum_{p\geq1}\e^{p-1}\F^o_p.
\eeq
\end{lem}
\begin{prf}
By using~\eqref{Virasoroconstraints} and~\eqref{generateL},
we have
\beq\label{Ltau}
\frac1{\tau_{\rm top}}\mathcal L(\lambda)\left(\tau_{\rm top}\right)=0.
\eeq
By substituting~\eqref{lij} into the above~\eqref{Ltau}, and then taking the coefficients of $\e^{-2}$, $\e^{-1}$ and the remaining part,
we get~\eqref{JF1}, \eqref{JF2} and~\eqref{loop0} respectively.
Hence the lemma is proved.
\end{prf}

Now we are ready to prove Theorem~\ref{main1}.

\begin{prfn}{Theorem~\ref{main1}}
Let us denote
\beq
\D(\lambda):=
-\left[\left(\I_1(\lambda)\J_1(\lambda)+\I_2(\lambda)\J_2(\lambda)\right)\right]_{-}
-\J_1(\lambda)(\F^{\rm c}_0)\J_1(\lambda).
\eeq
Then equation~\eqref{loop0} can be written as follows:
\begin{align}
\D(\lambda)(\Delta\F)=&\frac{\e^2}2\left(\J_1(\lambda)^2(\Delta \F)+\left(\J_1(\lambda)(\Delta\F)\right)^2\right)
+\e\left(\J_1(\lambda)(\F^{\rm o}_0)\J_1(\lambda)(\Delta\F)-\frac34\p_\lambda \J_2(\lambda)(\Delta\F)\right)\nn\\
& +\frac12\J_1(\lambda)^2(\F^{\rm c}_0)+\frac{\e}{2}\J_1(\lambda)^2(\F^{\rm o}_0)
+\frac12\J_1(\lambda)(\F^{\rm o}_0)^2-\frac34\p_\lambda \J_2(\lambda)(\F^{\rm o}_0)+\frac{5}{8\lambda^2}.
\label{loop1}
\end{align}

Recall that $v=\p_x^2\F^c_0$ and $r=\p_x\F^o_0$.
It follows from the genus-zero part of equations \eqref{openkdv1}--\eqref{openkdv2} that for $k\ge0$,
\begin{align}
&\frac{\p v}{\p t_k}=\frac{v^k}{k!}v_x,
\quad 
\frac{\p r}{\p t_k}=\sum_{i=0}^{k-1} \frac{v^ir^{2k-2i+1}}{i!(2k-2i+1)!!}(v_x+rr_x)+\frac{v^k}{k!}r_x,
\\
&\frac{\p v}{\p s_k}=0,\quad \frac{\p r}{\p s_k}=\frac1{k!}\left(v+\frac{r^2}2\right)^k(v_x+rr_x).
\end{align}
Then we have
\begin{align}
&\J_1(\lambda)(\p_x\F^{\rm c}_0)=
\sum_{i\geq0}\frac{(2i+1)!!}{2^{i+1}}\lambda^{-i-\frac32} \frac{v^{i+1}}{(i+1)!}
=A^{\frac12}-\lambda^{-\frac12}, \label{loopData1}\\
&\J_1(\lambda)^2(\F^{\rm c}_0)=
\sum_{i,j\geq0}\frac{(2i+1)!!(2j+1)!!}{2^{i+j+2}i!j!(i+j+1)}\lambda^{-i-j-3}v^{i+j+1}
=\frac1{8}A^2-\frac1{8}\lambda^{-2},\\
&\p_\lambda J_2(\lambda)(\F^{\rm o}_0)
=-\sum_{i\geq0}(i+2)\lambda^{-i-3}\big(v+\frac{1}2r^2\big)^{i+1}
=\lambda^{-2}-B^2,\\
&\J_1(\lambda)(\F^{\rm o}_0)=\sum_{i\geq0}\sum_{j=0}^i \frac{(2i+1)!!}{2^{i+1}}\lambda^{-i-\frac32}
\frac{v^j r^{2i-2j+1}}{j!(2i-2j+1)!!}
=\frac{r}{2}A^{\frac12}B,\\
&\J_1(\lambda)^2(\F^{\rm o}_0)=\sum_{i\geq0}\frac{(2i+1)!!}{2^{i+1}}\lambda^{-i-\frac32}
\frac{\p }{\p t_i} \mathcal{J}_1(\lambda)(\F^{\rm o}_0)
=\frac{A}{8r}\p_x\left(r^2 AB\right)+\frac{r}{6}\p_x\left(B^3\right). \label{loopData2}
\end{align}
Here we recall that $A$ and $B$ are given by \eqref{defab}.
It follows from~\eqref{JF1} and~\eqref{JF2} that
\beq
\mathcal D(\lambda)(v)=A,\quad
\mathcal D(\lambda)(r)=\frac{1}{2}r AB.
\eeq
Note that $\left[D(\lambda),~\p_x\right]=A^{\frac12}\J_1(\lambda)$.
Then we obtain by induction that
\begin{align}
&\mathcal D(\lambda)(v_i)
=\p_x^i(A)+\sum_{j=1}^i\binom i j\p_x^{j-1}\big(A^{\frac12}\big)\p_x^{i+1-j}\big(A^{\frac12}\big), \label{loopData3}\\
&\mathcal D(\lambda)(r_i)
=\frac12\p_x^i\big(r A B \big)
+\frac12\sum_{j=1}^i\binom i j\p_x^{j-1}\big( A^{\frac12} \big)\p_x^{i+1-j}\big(r A^{\frac12} B\big). \label{loopData4}
\end{align}	

Using~\eqref{loopData1}--\eqref{loopData2}, \eqref{loopData3}, \eqref{loopData4} and \eqref{loop1}, we arrive at~\eqref{loopequation}. 
Here we note that equation~\eqref{loopequation} holds identically in $\lambda$.

It remains to show the uniqueness of the solution to equation~\eqref{loopequation}.
By using Theorem~\ref{main0}, we can write $\Delta F$ into the form
\beq\label{deltaf}
\Delta F=:\sum_{p\geq0}\e^{p} X_p(r,v_1,r_1,\dots,v_{2p+1},r_{2p+1}).
\eeq 
Hence, by taking the coefficient of $\e^p$ in~\eqref{loopequation},
we have 
\begin{align}
&\sum_{i=0}^{2p+1}\frac{\p X_p}{\p v_{i}}\bigg(\p^i (A)+\sum_{j=1}^i\binom i j\p^{j-1}\left(A^{\frac12}\right)\p^{i+1-j}\big(A^{\frac12}\big) \bigg)\nn\\
&+\frac12\sum_{i=0}^{2p+1}\frac{\p X_p}{\p r_{i}}
\bigg(\p^i\left(r A B\right)+\sum_{j=1}^i\binom i j\p^{j-1}\left(A^{\frac12}\right)\p^{i+1-j}
\left(r A^{\frac12} B\right) \bigg)\nn\\
=&\frac{1}{2}\sum_{i=0}^{2p-3}
\bigg(
\frac{\p X_{p-2}}{\p v_{i}}\p^{i+2}\Big(\frac{A^2}{8}\Big)
+\frac{\p X_{p-2}}{\p r_{i}}\p^{i+1}\Big(\frac{A}{8r}\p\big(r^2 A B\big)
+\frac{r}{6}\p\big(B^3\big)\Big)
\bigg)  \nn\\
&+\frac{1}{2}\sum_{i,j=0}^{2g-3}
\bigg[\bigg(\frac{\p^2 X_{p-2}}{\p v_i \p v_j}
+\sum_{k=0}^{p-2}\frac{\p X_k}{\p v_{i}}\frac{\p X_{p-k-2}}{\p v_{j}}\bigg)
\p^{i+1}\big(A^{\frac12}\big)\p^{j+1}\big(A^{\frac12}\big)  \nn\\
&+\frac14\bigg(\frac{\p^2 X_{p-2}}{\p r_{i} \p r_{j}}
+\sum_{k=0}^{p-2}\frac{\p X_k}{\p r_{i}}\frac{\p X_{p-k-2}}{\p r_{j}}\bigg)
\p^{i+1}\big(rA^{\frac12} B\big)\p^{j+1}\big(rA^{\frac12} B\big)  \nn\\
&+\bigg(\frac{\p^2 X_{p-2}}{\p v_{i} \p r_{j}}
+\sum_{k=2}^{p-2}\frac{\p X_k}{\p v_{i}}\frac{\p X_{p-k-2}}{\p r_{j}}\bigg)
\p^{i+1}\big(A^{\frac12}\big)\p^{j+1}\big(rA^{\frac12} B\big) \bigg]  \nn\\
&+\sum_{i=0}^{2p-1}\bigg(\frac12\frac{\p X_{p-1}}{\p v_{i}}
r A^{\frac12} B \p^{i+1}\big(A^{\frac12}\big)
+\frac{\p X_{p-1}}{\p r_{i}}
\Big(\frac14rA^{\frac12}B\p^{i+1}\big(rA^{\frac12}B\big)+\frac{3}{4}\p^{i+1}\big(B^2\big)\Big)\bigg)  \nn\\
&+\delta_{p,1}\Big(\frac{A}{16r}\p\left(r^2AB\right)+\frac{r}{12}\p\left(B^3\right)\Big)
+\delta_{p,0}\Big(\frac{A^2}{16}+B^2-\frac{1}{4}AB\Big)
\label{loopeq1}
\end{align}
By writing it into partial fraction with respect to $\lambda$, 
and then taking the coefficients of the powers of $A$ and $B$,
we arrive at a system of equations in the following form:
\beq\label{loopeq2}
M_p
\left(\frac{\p X_p}{\p v},\frac{\p X_p}{\p r},\dots \frac{\p X_p}{\p v_{2p+1}},\frac{\p X_p}{\p r_{2p+1}}\right)^T
=C_p,
\eeq
where $M_p$ is an upper triangular matrix with diagonal elements being
\beq
1,\frac1r, \dots, \frac{(2i+1)!!}{2^i}v_1^i, \frac{(i+1)!}{r}\left(v_1+rr_1\right)^i,\dots,\frac{(4p+3)!!}{2^{2p+1}}v_1^{2p+1},\frac{(2p+2)!}{r}\left(v_1+rr_1\right)^{2p+1},
\eeq
and $C_p$ is a column vector determined by $X_0,\dots,X_{p-1}$.
Then the gradient of~$X_p$ can be uniquely determined by the system of linear equations~\eqref{loopeq2}.
The theorem is proved.
\end{prfn}

Using the loop equation, we obtain the expressions in Example \ref{emp1} as well as
\begin{align*}
F^o_3=&
\frac{1}{v_1^4 \left(v_1+r r_1\right)^6}
\Big(
\frac{1}{8} r_2^2 v_1^8+\frac{5}{24} r_1 r_3 v_1^8+\frac{73 v_4 v_1^8}{1152}+\frac{1}{12} r r_4 v_1^8-\frac{1}{8} r r_1 r_2^2 v_1^7-\frac{9}{16} r_1^3 r_2 v_1^7
\\&
-\frac{7}{12} r_1 v_2 r_2 v_1^7-\frac{3}{16} r_1^2 v_3 v_1^7-\frac{149}{640} v_2 v_3 v_1^7-\frac{13}{48} r r_2 v_3 v_1^7+\frac{35}{48} r r_1^2 r_3 v_1^7-\frac{7}{24} r v_2 r_3 v_1^7
\\&
-\frac{1}{8} r^2 r_2 r_3 v_1^7+\frac{19}{64} r r_1 v_4 v_1^7+\frac{23}{48} r^2 r_1 r_4 v_1^7+\frac{1}{48} r^2 v_5 v_1^7+\frac{1}{48} r^3 r_5 v_1^7+\frac{1}{8} r_1^6 v_1^6+\frac{61}{360} v_2^3 v_1^6
\\&
-\frac{1}{4} r^3 r_2^3 v_1^6+\frac{5}{12} r_1^2 v_2^2 v_1^6-\frac{23}{8} r^2 r_1^2 r_2^2 v_1^6+\frac{1}{6} r^2 v_2 r_2^2 v_1^6-\frac{1}{8} r^2 v_3^2 v_1^6-\frac{1}{8} r^4 r_3^2 v_1^6+\frac{19}{48} r_1^4 v_2 v_1^6
\\&
-\frac{31}{16} r r_1^4 r_2 v_1^6+\frac{7}{12} r v_2^2 r_2 v_1^6-\frac{67}{48} r r_1^2 v_2 r_2 v_1^6-\frac{19}{24} r r_1^3 v_3 v_1^6-\frac{307}{320} r r_1 v_2 v_3 v_1^6-\frac{29}{16} r^2 r_1 r_2 v_3 v_1^6
\\&
+\frac{1}{48} r^2 r_1^3 r_3 v_1^6-\frac{43}{24} r^2 r_1 v_2 r_3 v_1^6-\frac{49}{24} r^3 r_1 r_2 r_3 v_1^6-\frac{1}{4} r^3 v_3 r_3 v_1^6+\frac{133}{384} r^2 r_1^2 v_4 v_1^6-\frac{1}{6} r^2 v_2 v_4 v_1^6
\\&
-\frac{1}{6} r^3 r_2 v_4 v_1^6+\frac{37}{48} r^3 r_1^2 r_4 v_1^6-\frac{1}{6} r^3 v_2 r_4 v_1^6-\frac{1}{6} r^4 r_2 r_4 v_1^6+\frac{1}{16} r^3 r_1 v_5 v_1^6+\frac{1}{16} r^4 r_1 r_5 v_1^6
\\&
+\frac{1}{2} r r_1^7 v_1^5+\frac{3}{5} r r_1 v_2^3 v_1^5+2 r^4 r_1 r_2^3 v_1^5+\frac{73}{48} r r_1^3 v_2^2 v_1^5+\frac{3}{8} r^3 r_1^3 r_2^2 v_1^5+\frac{67}{12} r^3 r_1 v_2 r_2^2 v_1^5-\frac{1}{4} r^3 r_1 v_3^2 v_1^5
\\&
-\frac{1}{4} r^5 r_1 r_3^2 v_1^5+\frac{73}{48} r r_1^5 v_2 v_1^5+\frac{13}{16} r^2 r_1^5 r_2 v_1^5+\frac{25}{6} r^2 r_1 v_2^2 r_2 v_1^5+\frac{239}{48} r^2 r_1^3 v_2 r_2 v_1^5-\frac{1}{4} r^2 r_1^4 v_3 v_1^5
\\&
+\frac{3}{4} r^2 v_2^2 v_3 v_1^5+\frac{3}{4} r^4 r_2^2 v_3 v_1^5+\frac{33}{128} r^2 r_1^2 v_2 v_3 v_1^5-\frac{21}{16} r^3 r_1^2 r_2 v_3 v_1^5+\frac{3}{2} r^3 v_2 r_2 v_3 v_1^5-\frac{9}{16} r^3 r_1^4 r_3 v_1^5
\\&
+\frac{3}{4} r^3 v_2^2 r_3 v_1^5+\frac{3}{4} r^5 r_2^2 r_3 v_1^5-\frac{29}{24} r^3 r_1^2 v_2 r_3 v_1^5-\frac{53}{24} r^4 r_1^2 r_2 r_3 v_1^5+\frac{3}{2} r^4 v_2 r_2 r_3 v_1^5-\frac{1}{2} r^4 r_1 v_3 r_3 v_1^5
\\&
+\frac{17}{288} r^3 r_1^3 v_4 v_1^5-\frac{1}{3} r^3 r_1 v_2 v_4 v_1^5-\frac{1}{3} r^4 r_1 r_2 v_4 v_1^5+\frac{7}{16} r^4 r_1^3 r_4 v_1^5-\frac{1}{3} r^4 r_1 v_2 r_4 v_1^5-\frac{1}{3} r^5 r_1 r_2 r_4 v_1^5
\\&
+\frac{1}{16} r^4 r_1^2 v_5 v_1^5+\frac{1}{16} r^5 r_1^2 r_5 v_1^5-\frac{1}{8} r^2 r_1^8 v_1^4-\frac{1}{2} r^2 v_2^4 v_1^4-\frac{1}{2} r^6 r_2^4 v_1^4-\frac{37}{24} r^2 r_1^2 v_2^3 v_1^4+\frac{1}{4} r^5 r_1^2 r_2^3 v_1^4
\\&
-2 r^5 v_2 r_2^3 v_1^4-\frac{95}{48} r^2 r_1^4 v_2^2 v_1^4-3 r^4 v_2^2 r_2^2 v_1^4-\frac{7}{12} r^4 r_1^2 v_2 r_2^2 v_1^4-\frac{1}{8} r^4 r_1^2 v_3^2 v_1^4-\frac{1}{8} r^6 r_1^2 r_3^2 v_1^4
\\&
-\frac{43}{48} r^2 r_1^6 v_2 v_1^4+\frac{3}{16} r^3 r_1^6 r_2 v_1^4-2 r^3 v_2^3 r_2 v_1^4-\frac{29}{12} r^3 r_1^2 v_2^2 r_2 v_1^4-\frac{3}{16} r^3 r_1^4 v_2 r_2 v_1^4+\frac{3}{8} r^3 r_1^5 v_3 v_1^4
\\&
+\frac{3}{4} r^3 r_1 v_2^2 v_3 v_1^4+\frac{3}{4} r^5 r_1 r_2^2 v_3 v_1^4+\frac{91}{96} r^3 r_1^3 v_2 v_3 v_1^4+\frac{11}{48} r^4 r_1^3 r_2 v_3 v_1^4+\frac{3}{2} r^4 r_1 v_2 r_2 v_3 v_1^4-\frac{1}{16} r^4 r_1^5 r_3 v_1^4
\\&
+\frac{3}{4} r^4 r_1 v_2^2 r_3 v_1^4+\frac{3}{4} r^6 r_1 r_2^2 r_3 v_1^4+\frac{7}{24} r^4 r_1^3 v_2 r_3 v_1^4-\frac{7}{24} r^5 r_1^3 r_2 r_3 v_1^4+\frac{3}{2} r^5 r_1 v_2 r_2 r_3 v_1^4-\frac{1}{4} r^5 r_1^2 v_3 r_3 v_1^4
\\&
-\frac{19}{384} r^4 r_1^4 v_4 v_1^4-\frac{1}{6} r^4 r_1^2 v_2 v_4 v_1^4-\frac{1}{6} r^5 r_1^2 r_2 v_4 v_1^4+\frac{1}{16} r^5 r_1^4 r_4 v_1^4-\frac{1}{6} r^5 r_1^2 v_2 r_4 v_1^4-\frac{1}{6} r^6 r_1^2 r_2 r_4 v_1^4
\\&
+\frac{1}{48} r^5 r_1^3 v_5 v_1^4+\frac{1}{48} r^6 r_1^3 r_5 v_1^4+\frac{1}{18} r^3 r_1^3 v_2^3 v_1^3-\frac{5}{48} r^3 r_1^5 v_2^2 v_1^3-\frac{1}{48} r^3 r_1^7 v_2 v_1^3+\frac{1}{48} r^4 r_1^5 v_2 r_2 v_1^3
\\&
+\frac{1}{48} r^4 r_1^6 v_3 v_1^3-\frac{7}{128} r^4 r_1^4 v_2 v_3 v_1^3+\frac{1}{192} r^5 r_1^5 v_4 v_1^3+\frac{1}{24} r^4 r_1^4 v_2^3 v_1^2-\frac{1}{48} r^4 r_1^6 v_2^2 v_1^2-\frac{7}{320} r^5 r_1^5 v_2 v_3 v_1^2
\\&
+\frac{r^6 r_1^6 v_4 v_1^2}{1152}+\frac{1}{60} r^5 r_1^5 v_2^3 v_1-\frac{7 r^6 r_1^6 v_2 v_3 v_1}{1920}+\frac{1}{360} r^6 r_1^6 v_2^3
\Big)
\nn\\
=&
-\frac{r^2 u_2^4}{u_1^6}+\frac{r u_2^2(r_1 u_2+3r u_3)}{4u_1^5}
-\frac{r_1^2u_2^2-4 u_2^3-2 r r_2 u_2^2+6rr_1u_2u_3+3r^2u_3^2+4r^2u_2u_4 }{24u_1^4} \\
&
+\frac{r_1^2(r_1^2 u_2-u_2^2)}{48 v_1 u_1^3}
-\frac{r_1^2(r_1^4-2r_1u_2+u_2^2+rr_2u_2)}{48v_1^2 u_1^2}
-\frac{3r_1^3r_2-r_1^2 u_3}{48v_1u_1^2}
-\frac{2r_1^3r_2+3r_1r_2u_2}{48v_1^2u_1} \\
&
+\frac{2r_1r_2u_2+r_2^2u_3-11u_2u_3-2rr_2u_3-3rr_3u_2+2rr_1u_4+r^2 u_5}{48u_1^3} \\
&
-\frac{2r_2^2+r_1r_3-3u_4-rr_4}{48u_1^2}
-\frac{r_1r_3}{48 v_1u_1}
-\frac{r_2^2}{48v_1^2}.
\end{align*}
We observe that the expressions of $F_p^o$ in terms of $r,u_k,r_k$ are much shorter 
than those in terms of $r,v_k,r_k$.

\section{Applications}\label{sec4}

In this section, we give some applications of Theorems \ref{main0} and \ref{main1}.
Up to now, by solving the loop equation \eqref{loopeq1} and by using the expressions \eqref{Fc12}, 
we have obtained the expressions of the first two open free energies $F^o_p$ (see Example~\ref{emp1}).
Next we will further study how to relate our results and observations with the Taylor coefficients in the power series $\mathcal F^o_p$.

Let us consider the change of the independent variables in $F_p^o,p\geq1$ from the  jet variables to the Itzykson-Zuber type variables, as the latter have more direct relations with the correlators.
From the recursion relations~\eqref{Ijet1}--\eqref{Ijet2}, we obtain 
\begin{align}
&\sum_{k\geq0} k v_k\frac{\p I_n}{\p v_k}=\delta_{n,1}-I_n,
\quad \sum_{k\geq0} (k-1) v_k\frac{\p I_n}{\p v_k}=(n-1)I_n,
\label{IJcond1}\\
&\sum_{k\geq0} \Big(k v_k\frac{\p J_n}{\p v_k}+k r_k \frac{\p J_n}{\p r_k}\Big) =-J_n,
\quad
\sum_{k\geq0} \Big((k-1) v_k\frac{\p J_n}{\p v_k}+\Big(k-\frac12\Big) r_k \frac{\p J_n}{\p r_k}\Big) =\Big(n-\frac12\Big)J_n,
\label{IJcond2}
\end{align}
and by using the formula
\beq
\frac{\p}{\p x}=\frac1{1-I_1}\frac{\p}{\p I_0}+\sum_{k\geq1}\frac{I_{k+1}}{1-I_1}\frac{\p}{\p I_k}+\sum_{k\geq0}\frac{J_{k+1}+\frac{J_1}{1-I_1} I_{k+1}}{1-I_1-J_0J_1}\frac{\p}{\p J_k},
\eeq
we have the form:
\begin{align}
&v_k=\sum_{\substack{\lambda\in\mathcal{P}_{k-1}}}
C'_{k;\lambda} \frac{I_{\lambda+1}}{U_1^{k+\ell(\lambda)}},
\\
&r_k=\sum_{d=0}^{k+1}\sum_{\substack{\lambda,\mu\in\mathcal{P} \\ |\lambda|+|\mu|=k-\frac{-\ell(\mu)-d+1}2}} 
C''_{k;d;\lambda;\mu}(U_1^{-1}, U_2^{-1}) 
J_0^d I_{\lambda+1} J_{\mu},
\end{align}
where $C'_{k;\lambda}$ are certain rational numbers, and
$C''_{k;d;\lambda;\mu}(U_1^{-1}, U_2^{-1})$ are homogeneous polynomials in $U_1^{-1}$ and $U_2^{-1}$
of degree $m+n+k$ with rational coefficients.
Then by using Theorem \ref{main0}, we have that
\beq
F^o_1=-\frac12\log\left(1-I_1-J_0 J_1\right),
\eeq
and the function $F^o_p,\,p\geq2$ can be expressed as rational functions in $J_0,I_1,J_1,\dots,I_{2p-1},J_{2p-1}$ satisfying the conditions
\begin{align}
&-\sum_{k\geq1}\left(\left(I_k-\delta_{k,1}\right)\frac{\p F^o_p}{\p I_k}+J_k\frac{\p F^o_p}{\p J_k}\right)=(p-1)F_p^o,
\label{FopIJ1}
\\
&\sum_{k\geq1}\left((k-1)I_k\frac{\p F^o_p}{\p I_k}+\left(k-\frac12\right)J_k\frac{\p F^o_p}{\p J_k}\right)=\frac32(p-1)F_p^o.
\label{FopIJ2}
\end{align} 
For example,
\begin{align}
F^o_2=&
\frac{1}{\left(1-I_1\right)^2 \left(1-I_1-J_0J_1\right)^3}
\bigg(
\frac{1}{24} \left(3 J_3 J_0^2+3 I_3 J_0+4 J_2\right) (I_1-1)^3
\nn \\&
+\frac{1}{24} \left(-5 J_2^2 J_0^3+3 J_1 J_3 J_0^3+3 I_3 J_1 J_0^2-10 I_2 J_2 J_0^2-5 I_2^2 J_0-6 J_1 J_2 J_0-12 I_2 J_1\right) (I_1-1)^2
\nn \\&
+\left(\frac{5}{24} J_1^3-\frac{1}{8} J_0 J_1^2 I_2\right) (I_1-1)-\frac{1}{24} J_0^2 J_1^3 I_2
\bigg). \label{Fo2IJ}
\end{align}
In general, $F^o_p,\, p\geq2$ can be written into the form
\beq\label{exprFop}
F^o_p=
\frac{
\sum_{\substack{m,n\geq0,\\ m+n\leq 13p-7}}\sum_{\substack{2\leq k_1,\dots,k_m\leq2p-1\\ 1\leq \ell_1,\dots,\ell_n\leq 2p-1}}
\frac{B_{k_1,\dots,k_m;\ell_1,\dots,\ell_n}}{m!n!} \prod_{i=1}^m \frac{J_0^{2k_i-2}I_{k_i}}{1-I_1}\prod_{i=1}^n \frac{J_0^{2\ell_i-1}J_{\ell_i}}{1-I_1}
}
{J_0^{3p-3}(1-I_1)^{p-1}(1-\frac{J_0J_1}{1-I_1})^{7p-4}}. 
\eeq

Similar to~\cite{DY}, we consider the restriction of $\mathcal F^o_p({\bf t},{\bf s}),\,p\geq1$ at
\beq\label{tsrestriction}
t_0=t_1=0, \quad \sum_{k\geq 2}\frac{t_{k}}{(2k+1)!}+\sum_{k\geq0} \frac{s_k}{2^k k!}=1.
\eeq
According to the formulae \eqref{g0elkdv}, \eqref{g0el} and Lemma \ref{lemst}, this is equivalent to consider the restriction of $F^o_p,\,p\geq1$ at 
$I_0=I_1=0,\, J_0=1$ as follows:
\begin{align}
&F^{o,*}_1:=F_1^o\big|_{I_1=0,J_0=1}=-\frac12\log(1-J_1),\\
&F^{o,*}_p
:= F_p^o\big|_{I_1=0,J_0=1}
\in (1-J_1)^{-(7p-4)}\CC[I_2,\dots,I_{2p-1},J_1,\dots,J_{2p-1}],\quad p\geq2.
\label{fgostar}
\end{align}
(Here since $F^o_p$, $p\geq1$ are independent of $I_0$,  we do not write $I_0=0$ in the above formulae.)
The expression of $F^{o,*}_p$ is simpler  than $F^o_p$.
For $p\geq2$, we can recover all the information of $F^o_p$ from $F^{o,*}_p$ by 
\beq
F^o_p=\frac1{J_0^{3p-3}(1-I_1)^{p-1}}\left. F^{o,*}_p \right|_{I_k\rightarrow (1-I_1)^{-1}J_0^{2k-2} I_k, \, J_k\rightarrow (1-I_1)^{-1}J_0^{2k-1} J_k}.
\eeq
In other words, the restriction of $\mathcal F^o_p({\bf t},{\bf s})$, $p\geq2$ at \eqref{tsrestriction} has no loss of information.
Below we will use the expressions of $F^{o,*}_p$ to give relations among the correlators.
Indeed, we have, by using Lemma~\ref{lemst},
that
\beq\label{fofo}
F^{o,*}_p
=\mathcal F^o_p\left(t_0^*,s_0^*,t_1^*,s_1^*,\dots\right), \quad p\geq2
\eeq
where 
\begin{align}
&t_0^*=0,
\quad
s_0^*=1+\sum_{n\geq1}\frac{(-1)^n}{2^n n!} \left(J_{n}-\frac{1}{2n+1}I_{n+1}\right),
\label{tstar1}\\
&t_k^*=(1-\delta_{k,1})I_k,
\quad 
s_k^*=\sum_{n\geq0}\frac{(-1)^n}{2^n n!} \left(J_{k+n}-\frac{1}{2n+1}I_{n+k+1}\right),
\quad 
k\geq1.\label{tstar2}
\end{align}

It is obvious from \eqref{exprFop} that $F^{o,*}_p$ has the form
\beq\label{exprFopstar}
F^{o,*}_p=\frac{1}{(1-J_1)^{7p-4}}\sum_{\substack{m,n\geq0,\\ m+n\leq 13p-7}} 
\frac{B_{k_1,\dots,k_m;\ell_1,\dots,\ell_n}}{m!n!} I_{k_1}\cdots I_{k_m} J_{\ell_1}\cdots J_{\ell_n}.
\eeq
For example,
\begin{align}
F^{o,*}_2=&
\frac{1}{24} (-I_2-5)-\frac{5}{8 \left(J_1-1\right)}+\frac{-3 I_2+I_3-2 J_2+J_3-5}{8 \left(J_1-1\right)^2}
\nn \\
&-\frac{5 \left(I_2^2+2 J_2 I_2+2 I_2+J_2^2+2 J_2+1\right)}{24 \left(J_1-1\right)^3}. \label{empFo2}
\end{align}

\begin{prfn}{Corollary \ref{cor1}}
By using Lemma~\ref{lemst}, we know that 
\begin{align}
&\left.t_0\right|_{I_0=I_1=0,\,J_0=1}=0,
\quad
\left.t_k\right|_{I_0=I_1=0,\,J_0=1}=(1-\delta_{k,1})I_k,
\\
&\left.s_0\right|_{I_0=I_1=0,\,J_0=1}=1+\sum_{n\geq1}\frac{(-1)^n}{2^n n!} \left(J_{n}-\frac{1}{2n+1}I_{n+1}\right),
\\
&\left.s_k\right|_{I_0=I_1=0,\,J_0=1}=\sum_{n\geq0}\frac{(-1)^n}{2^n n!} \left(J_{k+n}-\frac{1}{2n+1}I_{n+k+1}\right),
\end{align}
where $k\ge 1$.
Hence by taking the derivatives of~\eqref{fofo}, 
we obtain 
\beq\label{corid1}
\frac{\p^{m+n} F^{o,*}_p}{\p I_{k_1}\cdots \p I_{k_m} \p J_{\ell_1}\cdots \p J_{\ell_n}}
\Big|_{I_{\geq2}=J_{\geq1}=0}
=
\Lambda_{k_1,\dots, k_m;\ell_1,\dots,\ell_n}
(\mathcal F_p^o)
\big|_{s_0=1,t_{\geq 0}=s_{\geq1}=0},
\eeq
where $k_1,\dots,k_m\geq2$ and $\ell_1,\dots,\ell_n\geq1$ and the operator $\Lambda_{k_1,\dots, k_m;\ell_1,\dots,\ell_n}$ is defined by \eqref{defLambda}.
Then the corollary is proved.
\end{prfn}

The property \eqref{tautopcond} implies that
\beq
\langle \tau_{k_1}\cdots\tau_{k_m}\sigma_{\ell_1}\cdots \sigma_{\ell_n}\rangle_p^o\neq 0,
\eeq
only if 
\beq
k_1+\cdots+k_m+\ell_1+\cdots+\ell_n=\frac{3p-3}{2}+m+\frac{n}2.
\eeq
It follows that
\begin{align}
&\left.
\frac{\p^{m+n} \mathcal F^o_p}{\p t_{k_1}\cdots\p t_{k_m} \p s_{\ell_1}\cdots \p s_{\ell_n}}
\right|_{s_0=1,t_{\geq0}=s_{\geq1}=0}
=&\frac{\langle \sigma_0^{3-3p} \prod_{i=1}^m \big(\sigma_0^{2k_i-2}\tau_{k_i}\big)
\prod_{j=1}^n \big(\sigma_0^{2\ell_j-1}\sigma_{\ell_j}\big)\rangle_p^o}{\big(2\sum_{i=1}^m k_m+2\sum_{j=1}^n\ell_j-3(p-1)-2m-n\big)!} 
\label{corid2}
\end{align}
if 
\beq
k_1+\cdots+k_m+\ell_1+\cdots+\ell_n\geq \frac32(p-1)+m+\frac n2,
\eeq
and vanishes otherwise.
Hence the left-hand side of \eqref{exprFostarp} is just a linear combination of the correlators.
\begin{emp}
It follows from Corollary \ref{cor1} together with the formulae \eqref{corid1} and \eqref{corid2} that when $n$ is sufficiently large,
we have
\begin{align*}
B^{[p]}_{\varnothing;1^n} 
=&
(7p-4)!\sum_{m=0}^n\sum_{k=3p-3}^{n-m}\binom{n}{m}\binom{n-m}{k}\frac{(-1)^{n-k}}{2^{n-m-k}(7p-4-m)!}
\frac{\langle\sigma_0^{k-3p+3}\sigma_1^k\rangle_2^o}{(2k-n+m-3p+3)!}=0,
\\
B^{[p]}_{2;1^n}
=&(7p-4)!\sum_{m=0}^n\sum_{k=3p-3}^{n-m}\binom{n}{m}\binom{n-m}{k}\frac{(-1)^{n-k}}{2^{n-m-k}(7p-4-m)!}
\\
&\times
\left(\frac{\langle\tau_2\sigma_0^{k-3p+5}\sigma_1^k\rangle_p^o}{(2k-n+m-3p+6)!}
-\frac{\langle\sigma_0^{k-3p+4}\sigma_1^{k+1}\rangle_p^o}{(2k-n+m-3p+4)!}
+\frac16\frac{\langle\sigma_0^{k-3p+3}\sigma_1^k\rangle_p^o}{(2k-n+m-3p+2)!}
\right)=0.
\\
\end{align*}
\end{emp}

We also have a conjectural form for $F^{o}_p,\,p\geq2$:
\beq\label{exprFopCnj}
F^o_p=
\frac{\sum_{\substack{m,n\geq0\\0\leq m+n\leq 6p-6}} \sum_{\substack{2\leq k_1,\dots,k_m\leq2p-1\\ 1\leq \ell_1,\dots,\ell_n\leq 2p-1}} 
\frac{\widetilde B^{[p]}_{k_1,\dots,k_m;\ell_1,\dots,\ell_n}}{m!n!}
\prod_{i=1}^m \frac{J_0^{2k_i-2} I_{k_i}}{1-I_1} \prod_{i=1}^n \frac{J_0^{2\ell_i-1}J_{\ell_i}}{1-I_1}}
{J_0^{3p-3}(1-I_1)^{p-1}(1-\frac{J_0J_1}{1-I_1})^{3p-3}}.
\eeq
It would be interesting if the two conjectural forms~\eqref{cnjFo} and~\eqref{exprFopCnj} could be
proved to be equivalent using the transformation between $v_k, r_k$ and $I_k, J_k$.
Under the restriction $I_1=0,J_0=1$, the conjectural formula \eqref{exprFopCnj} becomes
\beq\label{exprFopstarCnj}
F^{o,*}_p=\frac{1}{(1-J_1)^{3p-3}}
\sum_{0\leq m+n\leq 6p-6} \widetilde B^{[p]}_{k_1,\dots,k_m;\ell_1,\dots,\ell_n} I_{k_1}\cdots I_{k_m} J_{\ell_1}\cdots J_{\ell_n},
\eeq
which is equivalent to~\eqref{conjecturalLambda}. 

\begin{emp}
For $p=2$, we have the explicit expression \eqref{empFo2} of $F_2^{o,*}$,
which implies  that the conjectural relations~\eqref{conjecturalLambda} hold true for $p=2$. 
For examples,
\begin{align*}
\widetilde B^{[2]}_{\varnothing;1^n} 
=&6\sum_{m=0}^3\sum_{k=3}^{n-m}\binom{n}{m}\binom{n-m}{k}\frac{(-1)^{n-k}}{2^{n-m-k}(3-m)!}
\frac{\langle\sigma_0^{k-3}\sigma_1^k\rangle_2^o}{(2k-n+m-3)!}=0
\end{align*}
for $n \ge 7$, and
\begin{align*}
\widetilde B^{[2]}_{2;1^n}
=&6\sum_{m=0}^3\sum_{k=3}^{n-m}\binom{n}{m}\binom{n-m}{k}\frac{(-1)^{n-k}}{2^{n-m-k}(3-m)!}
\\
&\times
\left(\frac{\langle\tau_2\sigma_0^{k-1}\sigma_1^k\rangle_2^o}{(2k-n+m-1)!}
-\frac{\langle\sigma_0^{k-2}\sigma_1^{k+1}\rangle_2^o}{(2k-n+m-2)!}
+\frac16\frac{\langle\sigma_0^{k-3}\sigma_1^k\rangle_2^o}{(2k-n+m-4)!}
\right)=0
\end{align*}
for $n\ge6$. 
But in fact, from the expression \eqref{empFo2}
we obtain more vanishing relations, namely, the above two identities hold for $n\ge 4$.

Since the explicit expression \eqref{empFo2} of $F_2^{o,*}$ is obtained,
we have even more relations:
\begin{align*}
&\widetilde B^{[2]}_{2,2;\varnothing}
=-2\langle\tau_2\sigma_1\rangle_2^o+\langle\tau_2^2\sigma_0\rangle_2^o=\frac5{12},
\\
&\widetilde B^{[2]}_{\varnothing;1,2}=-3\langle\sigma_2\rangle_2^o
+\langle\sigma_0\sigma_1\sigma_2\rangle_2^o
=\frac14,
\\
&\widetilde B^{[2]}_{3;\varnothing}=-\langle\sigma_2\rangle_2^o
+\langle\tau_3\sigma_0\rangle_2^o
=\frac18,
\\
&\widetilde B^{[2]}_{\varnothing;3}=-\frac12\langle\sigma_2\rangle_2^o
+\frac12\langle\sigma_0^2\sigma_3\rangle_2^o
=\frac18,
\\
&\widetilde B^{[2]}_{2;1,1}=\langle\tau_2\sigma_0\sigma_1^2\rangle_2^o
-\langle\sigma_1^3\rangle_2^o
-6\langle \tau_2\tau_1\rangle_2^o
=-\frac14,
\\
&\widetilde B^{[2]}_{\varnothing;2,2}=-\langle\sigma_0\sigma_1\sigma_2\rangle_2^o
+\frac16\langle\sigma_0^3\sigma_2^2\rangle_2^o
=\frac{5}{12},
\\
&\widetilde B^{[2]}_{2;1,1,1}
=9\langle\sigma_1^3\rangle_2^o-\langle\sigma_0\sigma_1^4\rangle_2^o+18\langle\tau_2\sigma_1\rangle_2^o-\frac{21}2\langle\tau_2\sigma_0\sigma_1^2\rangle_2^o+\frac12\langle\tau_2\sigma_0^2\sigma_1^3
=\frac1{4},
\\
&\widetilde B^{[2]}_{2;2}
=-\langle\sigma_0\sigma_1\sigma_2\rangle_2^o-\frac12\langle\tau_2\sigma_1\rangle_2^o+\frac12\langle\tau_2\sigma_0^2\sigma_2\rangle_2^o
=\frac5{12},
\\
&\widetilde B^{[2]}_{3;1}=3\langle\sigma_2\rangle_2^o-\langle\sigma_0\sigma_1\sigma_2\rangle_2^o-\frac72 \langle\tau_3\sigma_0\rangle_2^o+\frac12\langle\tau_3\sigma_0^2\sigma_1\rangle_2^o
=-\frac18,
\\
&\widetilde B^{[2]}_{\varnothing;1,3}=\frac32\langle\sigma_2\rangle_2^o
-\frac12\langle\sigma_0\sigma_1\sigma_2\rangle_2^o
-2\langle\sigma_0^2\sigma_3\rangle_2^o
+\frac16\langle\sigma_0^3\sigma_1\sigma_3\rangle_2^o
=-\frac18.
\end{align*}
\end{emp}

\section{Conclusions}
The genus zero part of the topological tau-function for the Burgers--KdV hierarchy evaluated at $t_1=t_2=\dots=0, s_1=s_2=\dots=0$ 
gives rise to 
\beq\label{specialcase}
F=\frac16 (v^1)^3,\quad F^o=v^1 v^2+\frac16(v^2)^3, 
\eeq
which satisfy the open WDVV equations~\cite{HS, PST}.
In general, starting with an $n$-dimensional Frobenius manifold, 
one can impose the open WDVV equations~\cite{HS}
to obtain an $(n+1)$-dimensional $F$-manifold (due to 
an observation from Alcolado\footnote{One of the authors D.Y. would like to thank Xujia Chen who kindly pointed out 
the interesting reference~\cite{A}.} and P.~Rossi).
A reconstruction of higher genus theory was given in~\cite{ABLR}  for flat F-manifolds.
It is also possible to reconstruct higher genus parts by using the Dubrovin--Zhang approach together with the results in \cite{BB} similarly to what  
we do in this paper.  
We plan to do general cases including extended $r$-spin~\cite{BCT0, BCT2} and the open analogue of a semisimple 
Frobenius manifold in future publications.

\begin{center}
{\small
Di Yang, School of Mathematical Sciences, University of Science and Technology of China,\\
Hefei 230026, P.R.~China\\
e-mail: diyang@ustc.edu.cn\\
~\\
Chunhui Zhou, Institute of Geometry and Physics, University of Science and Technology of China,\\ 
Hefei 230026, P.R.~China\\
e-mail: zhouch@ustc.edu.cn}
\end{center}


\begin{thebibliography}{}
\bibitem{A}
A. Alcolado. Extended Frobenius manifolds and the open WDVV equations. PhD Thesis. McGill University (Canada), 2017.

\bibitem{ABLR}
A. Arsie, A. Buryak, P. Lorenzoni, P. Rossi,
Semisimple flat F-manifolds in higher genus.
Comm. Math. Phys. {\bf 397} (2023), 141--197.

\bibitem{BB}
A. Basalaev, A. Buryak,
Open WDVV equations and Virasoro constraints.
Arnold Math. J. {\bf 5} (2019), 145--186.

\bibitem{BDY}
M. Bertola, B. Dubrovin, D. Yang,
Correlation functions of the KdV hierarchy and applications to intersection numbers over $\overline{\mathcal{M}}_{g,n}$.
Phys. D {\bf 327} (2016), 30--57.

\bibitem{BR}
M.~Bertola, G.~Ruzza, 
The Kontsevich-Penner matrix integral, isomonodromic tau functions and open intersection numbers. 
Ann. Henri Poincar\'e~{\bf 20} (2019), 393--443.

\bibitem{BY}
M.~Bertola, D.~Yang, 
The partition function of the extended $r$-reduced Kadomtsev-Petviashvili hierarchy. J. Phys. A~{\bf 48} (2015), Paper No.~195205, 20~pp.

\bibitem{BBCCN}
G. Borot, V. Bouchard, N. Chidambaram, T. Creutzig, D. Noshchenko, 
Higher Airy structures, $\mathcal{W}$-algebras and topological recursion.
Mem. Amer. Math. Soc.~{\bf 296} (2024), No.~{\bf 1476}, v+108 pp.

\bibitem{Bu1} 
A. Buryak, 
Equivalence of the open KdV and the open Virasoro equations for the moduli space of Riemann surfaces with boundary.
Lett. Math. Phys. {\bf 105} (2015), 1427--1448.

\bibitem{Bu2} 
A. Buryak, 
Open intersection numbers and the wave function of the KdV hierarchy.
Mosc. Math. J. {\bf 16} (2016), 27--44.

\bibitem{BCT0}
A. Buryak, E. Clader, R. J. Tessler, Closed extended $r$-spin theory and the 
Gelfand-Dickey wave function. J. Geom. Phys.~{\bf 137} (2019), 132--153.

\bibitem{BCT2}
A. Buryak, E. Clader, R. J. Tessler, Open $r$-spin theory III: A prediction for higher genus.
J. Geom. Phys. {\bf 192} (2023), 12~pp.

\bibitem{BT} 
A. Buryak, R. J. Tessler, R. J. Tessler, 
Matrix models and a proof of the open analog of Witten's conjecture.
Comm. Math. Phys.~{\bf 353} (2017), 1299--1328.

\bibitem{DM}
P. Deligne, D. Mumford,
The irreducibility of the space of curves of given genus. 
Publ. Math. I.H.E.S.~{\bf 36} (1969), 75--109.

\bibitem{DW}
R. Dijkgraaf, E. Witten, 
Developments in topological gravity.
Internat. J. Modern Phys. A {\bf 33} (2018), 63~pp.

\bibitem{Du}
B. Dubrovin,
Geometry of 2D topological field theories. 
In: Francaviglia, M., Greco, S. (eds.) ``Integrable Systems and Quantum Groups" (Montecatini Terme, 1993), 
Lecture Notes in Math., Vol.~{\bf 1620}, Springer, Berlin, 1996, pp.~120--348.

\bibitem{DY}
B. Dubrovin, D. Yang,
Remarks on intersection numbers and integrable hierarchies. I. Quasi-triviality.
Adv. Theor. Math. Phys. {\bf 24} (2020), 1055--1085.

\bibitem{DZ-norm}
B. Dubrovin, Y. Zhang,
Normal forms of hierarchies of integrable PDEs, Frobenius manifolds and Gromov--Witten invariants.
arXiv:math/0108160.

\bibitem{EYY}
T. Eguchi, Y. Yamada, S.-K. Yang, On the genus expansion in the topological string theory. Rev. Maht. Phys. {\bf 7} (1995), 279--309


\bibitem{HS}
A. Horev, J. P. Solomon,
The open Gromov--Witten--Welschinger theory of blow-ups of the projective plane.
arXiv:12104043.

\bibitem{IZ}
C. Itzykson, J.B. Zuber, Combinatorics of the modular group II the Kontsevich integrals. Internat. J. Modern
Phys. A, {\bf 23} (1992), 5661--5705.

\bibitem{K}
M. Kontsevich, 
Intersection theory on the moduli space of curves and the matrix Airy function.
Comm. Math. Phys.~{\bf 147} (1992), 1--23.

\bibitem{LYZZ}
S.-Q. Liu, D. Yang, Y. Zhang, C. Zhou, 
The Hodge-FVH correspondence.
J. Reine Angew. Math.~{\bf 775} (2021), 259--300.

\bibitem{PST}
R. Pandharipande, J. P. Solomon, R. J. Tessler,
Intersection theory on moduli of disks, open KdV and Virasoro.
arXiv:1409.2191.

\bibitem{OK} 
K. Okuyama, K. Sakai:
Genus expansion of open free energy in 2d topological gravity.
J. High Energy Phys.~{\bf 2021} (2021), Paper No. 217, 21~pp.

\bibitem{W}
E. Witten, 
Two-dimensional gravity and intersection theory on moduli space.
In "Surveys in differential
geometry (Cambridge, MA, 1990)", pp.~243--310. Lehigh Univ., Bethlehem, PA, 1991.

\bibitem{YZ}
D. Yang, C. Zhou,
On an extension of the generalized BGW tau-function.
Lett. Math. Phys.~{\bf 111} (2021), 23~pp.

\bibitem{Zhou}
J. Zhou, On topological 1D gravity. I. arXiv:1412.1604.

\end{thebibliography}
\end{document}